\documentclass[twocolumn, twocolappendix]{aastex631}
\usepackage{amsmath}
\usepackage{enumitem}
\usepackage{newtxtext,newtxmath}
\usepackage{bm}

\hypersetup{breaklinks,colorlinks,citecolor=blue,urlcolor=blue,linkcolor=red}

\newcommand{\uatentry}[2]{\href{http://astrothesaurus.org/uat/#2}{#1 (#2)}}
\newcommand{\Msun}{\,{\rm M}_\odot}

\defcitealias{ibata_charting_2024}{I24}
\defcitealias{chen_improved_2025}{C25}

\graphicspath{{./}{figures/}}

\shorttitle{\it Mass loss rate of GCs}
\shortauthors{\it Chen, Li, \& Gnedin}

\begin{document}

\title{\vspace{-6mm}\large Stellar streams reveal the mass loss of globular clusters}
\author[0000-0002-5970-2563]{Yingtian Chen}\thanks{\href{mailto: ybchen@umich.edu}{ybchen@umich.edu}}
\affiliation{Department of Astronomy, University of Michigan, Ann Arbor, MI 48109, USA}
\author[0000-0002-1253-2763]{Hui Li}\thanks{\href{mailto: hliastro@tsinghua.edu.cn}{hliastro@tsinghua.edu.cn}}
\affiliation{Department of Astronomy, Tsinghua Univeristy, Beijing 100084, China}
\author[0000-0001-9852-9954]{Oleg Y. Gnedin}
\affiliation{Department of Astronomy, University of Michigan, Ann Arbor, MI 48109, USA}

\begin{abstract}\noindent 
Globular cluster (GC) streams, debris of stars that tidally stripped from their progenitor GCs, have densities that correlate positively with the GC mass loss rate. In this work, we employ a novel particle spray algorithm that can accurately reproduce the morphology of streams of various orbital types, enabling us to uncover the relationship between the GC mass loss history and stream density profiles. Using recent discoveries of GC streams from \textit{Gaia} DR3, we present, for the first time, a catalog of directly observed mass loss rates for 12 Galactic GCs, ranging from 0.5 to 200~$\rm M_\odot\,Myr^{-1}$. By fitting power-law relations between mass loss rate and key GC properties, we identify positive correlations with GC mass and orbital frequency, consistent with the predictions from N-body simulations.
\end{abstract}

\keywords{\uatentry{Stellar streams}{2166}; \uatentry{Globular star clusters}{656}; \uatentry{Stellar dynamics}{1596}; \uatentry{Galaxy dynamics}{591}}

\section{Introduction}

Tidal dissolution of globular clusters (GCs) leads to significant mass loss depending on GC properties and their galactic environment \citep{gnedin_destruction_1997,prieto_dynamical_2008,renaud_evolution_2011,lamers_evolution_2013,renaud_star_2018,meng_tidal_2022,gieles_mass-loss_2023}. In particular, a power-law scaling between the GC's mass loss rate and mass, $\dot{M} \propto M_{\rm GC}^a$, plays an important role in transforming the initial cluster mass function \citep{portegies_zwart_young_2010,krumholz_star_2019} to the present-day log-normal shape \citep{okazaki_evolution_1995,fall_dynamical_2001}. To efficiently disrupt the overabundant low-mass GCs while preserving the high-mass population, the slope $a$ must be less than 1. 

However, only theoretical efforts have been made to study this scaling. N-body simulations predict a range of the slope $a$ from $-1$ to $1/3$ \citep{baumgardt_dynamical_2003,lamers_mass-loss_2010,lamers_evolution_2013,gieles_mass-loss_2023}. Additionally, sub-grid and semi-analytical models of GC formation employed in cosmological simulations also constrain the GC mass loss rate by matching model populations to observations \citep{li_star_2019,reina-campos_introducing_2022,chen_modeling_2022,chen_formation_2023,chen_catalogue_2024}. Nevertheless, these constraints are sensitive to the local environments of GCs, which can vary significantly across models. Moreover, there has been no direct observational measurements of mass loss rates to support these models.

Fortunately, the rapidly growing number of stellar streams detected in the Milky Way (MW) offers a unique opportunity to infer GC mass loss rates from observations. Thanks to the \textit{Gaia} mission \citep{gaia_collaboration_gaia_2016}, more than 100 streams have been identified \citep{bonaca_stellar_2025}, 16 of which are associated with known GCs by \citet[][hereafter \citetalias{ibata_charting_2024}]{ibata_charting_2024}. Stellar streams are debris of stars that tidally stripped from GCs, where energy exchange through two-body interactions gradually pushes these stars beyond the tidal radius \citep{binney_galactic_2008,weatherford_stellar_2023}. The density of these streams correlates positively with the GC mass loss rate, making them a useful observational probe for determining $\dot{M}$.

The relationship between stream density and $\dot{M}$ is complicated by the degeneracy with the stream width and length. Recent development of the improved mock stream generation algorithm by \citet[][hereafter \citetalias{chen_improved_2025}]{chen_improved_2025} has allowed us to reproduce the morphology of stellar streams with an accuracy of better than 10\%, while treating the mass loss rate as an independent parameter. This helps to break the degeneracy and place direct constraints on $\dot{M}$.

Here we present the first catalog of directly observed mass loss rates for 12 Galactic GCs. We develop a robust pipeline to obtain unbiased estimates of the mass and density for \citetalias{ibata_charting_2024} streams, accounting for missing faint stars below the detection limit and the spatial selection function along and perpendicular to the stream track. We compare these density estimates with mock streams generated using the \citetalias{chen_improved_2025} algorithm to calculate the mass loss rates. The detailed methodology is described in \S\ref{sec:method}. In \S\ref{sec:results}, we present our measurement of $\dot{M}$ and explore correlations between $\dot{M}$ and three key GC properties. We summarize our findings in \S\ref{sec:summary}.

\newpage
\section{Method}
\label{sec:method}

We aim to reproduce the observed mass density of stellar streams by generating mock streams with varying mass loss rates $\dot{M}$. The best-fit $\dot{M}$ then provides an estimate of the actual mass loss rate of the progenitor GC. We first describe our observational sample in \S\ref{sec:observational_sample}, then the mock stream generation algorithm in \S\ref{sec:modeling}, and our method for comparing the observed and mock densities in \S\ref{sec:mass_loss_rate}. Due to the incompleteness of the observational data, some stream stars may be missed if they are: 1) fainter than the detection limit, 2) far from the stream track, or 3) in regions with high contamination. We address these issues in \S\ref{sec:obs_mass} and \S\ref{sec:selection_function}.

\subsection{Observational sample}
\label{sec:observational_sample}

We use the \citetalias{ibata_charting_2024} catalog of stellar streams. These authors applied the \texttt{STREAMFINDER} algorithm \citep{malhan_streamfinder_2018,malhan_ghostly_2018,ibata_charting_2021} to \textit{Gaia} DR3, detecting 87 thin streams in the MW. A total of 24,540 stars with $G<20$ were identified above the 6-$\sigma$ likelihood threshold of being stream members. \textit{Gaia} DR3 provides near-complete sky coverage up to $G \approx 20$ \citep{cantat-gaudin_empirical_2023}. Here, we apply a conservative magnitude cut at $G_{\rm limit}=19$ to ensure even higher completeness, but any value in the range $G_{\rm limit}=18-20$ does not significantly impact the results of our analysis.

Streams are commonly described in a great circle frame $(\phi_1, \phi_2)$, where $\phi_1$ follows the stream track and $\phi_2$ is perpendicular to it. We derive this frame by rotating the equatorial coordinates to minimize the root mean square residual (RMSR) in the $\phi_2$ direction. The zero-point of the coordinate system is set by minimizing the RMSR along $\phi_1$.

Even in the great circle frame, streams can still appear curved. To correct for this, we further fit the stream track $\phi_{2, {\rm track}}(\phi_1)$ in the great circle frame using a quartic polynomial, minimizing the RMSR of the residual perpendicular coordinate $\hat{\phi}_2 \equiv \phi_2 - \phi_{2, {\rm track}}(\phi_1)$. We describe streams either in the $(\phi_1, \phi_2)$ system or the $(\phi_1, \hat{\phi}_2)$ system, as needed.

\subsection{Generating mock GC streams}
\label{sec:modeling}

\citetalias{ibata_charting_2024} associated 16 known GCs with streams in their catalog, including NGC 288, NGC 1261, NGC 1851, NGC 2298, NGC 2808, NGC 3201/Gj\"oll, M68/Fj\"oll, $\omega$Centauri ($\omega$Cen)/Fimbulthul, NGC 5466, Palomar 5 (Pal 5), M5, NGC 6101, M92, NGC 6397, NGC 7089, and NGC 7099. Throughout this work, we use the GC name to refer to both the GC and the associated stream. Using the cluster masses, positions, and velocities from the \citet{hilker_galactic_2019} catalog\footnote{\url{https://people.smp.uq.edu.au/HolgerBaumgardt/globular/}}, we apply particle spray algorithms to generate mock tidal streams originating from these GCs. For Pal 5, however, we use the positions and velocities from \citet{erkal_sharper_2017}, as these authors carefully selected phase-space coordinates to match the observed stream track.

We use the implementation of the \citetalias{chen_improved_2025} particle spray algorithm\footnote{Tutorials for this algorithm are available at \url{https://github.com/ybillchen/particle_spray} and are preserved on Zenodo at \citet{chen_ybillchenparticle_spray_2024}.} in the package \texttt{agama} \citep{vasiliev_agama_2019}. This algorithm initializes the positions and velocities of stream tracer particles from a multivariate Gaussian distribution, calibrated through N-body simulations of disrupting GCs. The \citetalias{chen_improved_2025} algorithm accurately reproduces the width and length of simulated streams across a wide range of cluster mass and orbital type. Furthermore, \citetalias{chen_improved_2025} demonstrated that the initial phase space distribution of stream particles is independent of the mass loss rate $\dot{M}$ of the progenitor GC. Thus, varying $\dot{M}$ only scales the stream number density without altering its morphology.

To generate mock streams, we first integrate the past orbit of each GC from the present position to lookback time $t_{\rm begin}$. Then, we integrate the GC orbit forward from $t_{\rm begin}$ to the present day, releasing tracer particles with a number loss rate $\dot{N}_{\rm tracer}$. Each tracer particle represents a stellar population with mass $m$, with the initial positions and velocities determined by \citetalias{chen_improved_2025}. The orbits of these particles are integrated in the combined potential of the Galaxy and the GC, the latter is approximated by a Plummer sphere with core radius of 4~pc. We have verified that the specific value of the core radius does not significantly affect the particle orbits as long as it is much smaller than the GC tidal radius. The distribution of tracers at the present day forms the mock stream.

We use the three-component \texttt{MWPotential2014} model by \citet{bovy_galpy_2015} as the background Galactic potential. \texttt{MWPotential2014} successfully reproduces the morphology of most streams except for NGC 1261, NGC 5466, and NGC 7099, whose mock stream tracks deviate by more than one stream width. Other potential models like \citet{mcmillan_mass_2017} similarly fail to match the morphology of these streams. This misalignment is likely because these GCs have highly eccentric orbits with $e\approx0.8$. Also, the pericenters of NGC 1261 and NGC 7099 are only around 1~kpc from the Galactic center. Accurately reproducing their morphology would require a more detailed Galactic potential model near the center, which is beyond the scope of this work. Therefore, we exclude these streams from our analysis.

The start time $t_{\rm begin}$ must be earlier than the duration of the detected stream, to fully cover its span. Therefore, we set $t_{\rm begin} = 8$~Gyr, which exceeds the duration of all observed streams (Appendix~\ref{sec:duration}). As we show in \S\ref{sec:results}, as long as $t_{\rm begin}$ is greater than the stream’s observed duration, varying $t_{\rm begin}$ does not affect the distribution of tracer particles within the detected region of the stream. 

We assume a uniform number loss rate over time, $\dot{N}_{\rm tracer} = \text{constant}$. However, the actual rate may vary significantly at different phases of the orbit due to the time-varying tidal field. We have also tested an alternative ``pulsing'' model, where particles are only released at pericenters. The two models produce similar density distributions, indicating fast mixing of stars in the stream. The two models produce similar density distributions, indicating fast mixing of stars in the stream. The two models represent the two most extreme cases, and the actual $\dot{N}_{\rm tracer}$ likely lies between them. However, the consistency between the two models suggests that our results are insensitive to the specific form of $\dot{N}_{\rm tracer}(t)$. The only exception is NGC 6101, where the GC is very close to the pericenter and likely has a high mass loss rate that creates overdensities near the GC. Since the GC has not yet reached the pericenter, the ``pulsing'' model fails to produce such overdensities. To avoid such complications, we use constant $\dot{N}_{\rm tracer}$ throughout this work. We emphasize that $\dot{N}_{\rm tracer}$ should be considered as an orbit-averaged rate over the stream’s duration.

\subsection{Calculating the mass loss rate}
\label{sec:mass_loss_rate}

Our mock streams are represented by tracer particles with mass $m$, which is a constant over time and space. Given the 2D number density of stream particles on the sky, $\Sigma_N(\phi_1, \hat{\phi}_2)$, we can calculate the mass density as $\Sigma(\phi_1, \hat{\phi}_2) = m\Sigma_N(\phi_1, \hat{\phi}_2)$. This relationship holds for any values of $\phi_1$ and $\hat{\phi}_2$. We can thus integrate it over the sky without breaking the equality,
\begin{align*}
    &\iint f(\phi_1,\hat{\phi}_2)\Sigma(\phi_1,\hat{\phi}_2) d\phi_1 d\hat{\phi}_2 \\
    =m&\iint f(\phi_1,\hat{\phi}_2)\Sigma_N(\phi_1,\hat{\phi}_2) d\phi_1 d\hat{\phi}_2
\end{align*}
where $f(\phi_1, \hat{\phi}_2)$ is an arbitrary weight function. Therefore, the  orbit-averaged mass loss rate can be written as:
\begin{equation}
    \dot{M} = \dot{N}_{\rm tracer} m = \dot{N}_{\rm tracer} \frac{\iint f(\phi_1,\hat{\phi}_2)\Sigma(\phi_1,\hat{\phi}_2) d\phi_1 d\hat{\phi}_2}{\iint f(\phi_1,\hat{\phi}_2)\Sigma_N(\phi_1,\hat{\phi}_2) d\phi_1 d\hat{\phi}_2}
    \label{eq:mass_loss_arbitrary_f}
\end{equation}

If we perform mock observations to the mock stream with a spatial selection function $f_{\rm sel}(\phi_1, \hat{\phi}_2)$, the total mass that is selected by this function is given by
\begin{equation}
    M_{\rm sel} = \iint f_{\rm sel}(\phi_1,\hat{\phi}_2)\Sigma(\phi_1,\hat{\phi}_2) d\phi_1 d\hat{\phi}_2
    \label{eq:m_obs}
\end{equation}
By substituting $f_{\rm sel}$ for $f$ in Eq.~(\ref{eq:mass_loss_arbitrary_f}), we can estimate the orbit-averaged mass loss rate as
\begin{equation}
    \dot{M}=\frac{\dot{N}_{\rm tracer}M_{\rm sel}}{\iint f_{\rm sel}(\phi_1,\hat{\phi}_2)\Sigma_N(\phi_1,\hat{\phi}_2) d\phi_1 d\hat{\phi}_2}
\end{equation}
where $\dot{N}_{\rm tracer}$ is a known input parameter from the particle spray method. Numerically, the denominator can be approximated using the importance sampling technique:
\begin{equation}
    \dot{M}\approx\frac{\dot{N}_{\rm tracer}M_{\rm sel}}{\sum_j f_{\rm sel}(\phi_{1,j},\hat{\phi}_{2,j})}
    \label{eq:mass_loss_exact}
\end{equation}
where the summation is over all tracer particles weighted by the spatial selection function. To compute the mass loss rate, we only need $M_{\rm sel}$ and $f_{\rm sel}$, which can be observationally obtained from the \citetalias{ibata_charting_2024} catalog.

For real observations, however, $M_{\rm sel}$ is not simply the sum of the masses of individual stars associated with the stream in the \citetalias{ibata_charting_2024} catalog, as it must account for stars below the detection limit. Additionally, $f_{\rm sel}(\phi_1, \hat{\phi}_2)$ is not a universal function due to the complexity of stream morphology and the specific configuration of \texttt{STREAMFINDER}. We detail our methods for estimating $M_{\rm sel}$ and $f_{\rm sel}(\phi_1, \hat{\phi}_2)$ in the following sections.

\subsection{Correcting for missing stars below the detection limit}
\label{sec:obs_mass}

To ensure high completeness, we only use stars with $G\leq19$ in the \citetalias{ibata_charting_2024} catalog. Therefore, it is necessary to correct for the stellar mass below the detection limit to obtain $M_{\rm sel}$. The mass of each star in a stream follows the present-day stellar mass function (MF), $\psi(m)$, for $m \leq m_{\rm max}$, where $m_{\rm max}$ is the maximum mass of surviving stars at the current cluster age. We adopt a power-law MF, $\psi(m) \propto m^{-\alpha}$, with a lower limit set to the hydrogen-burning threshold, $m_{\rm limit} = 0.08\ M_\odot$. Brown dwarfs below this limit are excluded, as their slope is poorly constrained in streams. Neglecting brown dwarfs has a minimal impact on the final results since they contribute less than a few percent to the total stellar mass.

The slope $\alpha$ for each stream is taken from the measurements by \citet{baumgardt_evidence_2023} for the progenitor GC. Typically, $\alpha$ ranges from $-0.5$ to $1.5$.

In a magnitude-limited survey, the minimum observable stellar mass $m_{\rm min}$ at a given distance $d$ can be determined by the isochrone of the stellar population. Given the mass–luminosity relation $m(L)$ of the isochrone, we can compute the minimum stellar mass as a function of distance: $m_{\rm min} = m(L_{\rm min}(d)) \equiv m_{\rm min}(d)$. We assume streams are single stellar populations using the same templates as in \citetalias{ibata_charting_2024} from the \texttt{PARSEC} isochrones \citep{bressan_span_2012}.

Similarly, we can compute the mass $m_i$ of each star in the stream using its apparent magnitude and distance $d_i$. Note that distances derived from \textit{Gaia} parallaxes often have large uncertainties. To obtain more accurate estimates, we fit the $d(\phi_1)$ relation from our mock streams as a quartic polynomial by minimizing the RMSR.

A star is not observable if $m_i < m_{\rm min}(d_i) \equiv m_{{\rm min},i}$. Simply summing the masses of observable stars underestimates the total stellar mass. We use the following expression to account for the missing stellar mass
\begin{equation}
    \tilde{M}_{\rm sel}\equiv\sum_{i=1}^{N_{\star}} \theta(m_i-m_{{\rm min},i}) \, m_i \, w_i\equiv\sum_{k=1}^{N_{\rm\star,obs}} m_k \, w_k
    \label{eq:estimate}
\end{equation}
where the first summation is over all stars, whereas the second is only over stars with $m_k>m_{{\rm min},k}$. Correspondingly, $N_{\star}$ represents the total number of stars while $N_{\rm\star,obs}$ represents the number of stars above the detection limit. Here, the Heaviside step function $\theta$ excludes unobservable stars, while the weight $w_i$ is a correction factor assigned to the $i$-th star
\begin{equation}
    w_i\equiv\frac{\int_{m_{\rm limit}}^{m_{\rm max}}m\psi(m)dm}{\int_{m_{{\rm min},i}}^{m_{\rm max}}m\psi(m)dm}.
    \label{eq:correction}
\end{equation}
Typically, $w_i$ ranges from 2 to 200 depending on the distance to the stream (Appendix~\ref{sec:mf}). In Appendix~\ref{sec:proof}, we prove that Eq.~(\ref{eq:estimate}) is an unbiased estimate of $M_{\rm sel}$. 

\citet{baumgardt_evidence_2023} also attempted to fit the MF with broken power-law functions, using a threshold at $0.4~\Msun$. We test this form of MF and find that the correction factors derived from the broken power-law MF are mostly consistent with those from the single-component power-law MF, with deviations less than $20\%$. This indicates that the specific shape of the MFs does not significantly influence our analysis.

It is worth noting that the MF in a stream may differ from that in the progenitor GC because less massive stars tend to be ejected first \citep{webb_variation_2021}. Obtaining the exact MF in streams is beyond the scope of this work. However, since most streams were released relatively recently ($<1$~Gyr, see Appendix~\ref{sec:duration}), we expect the two MFs to differ less significantly.

\subsection{Estimating the spatial selection function}
\label{sec:selection_function}

\begin{figure*}
    \centering
    \includegraphics[width=0.9\linewidth]{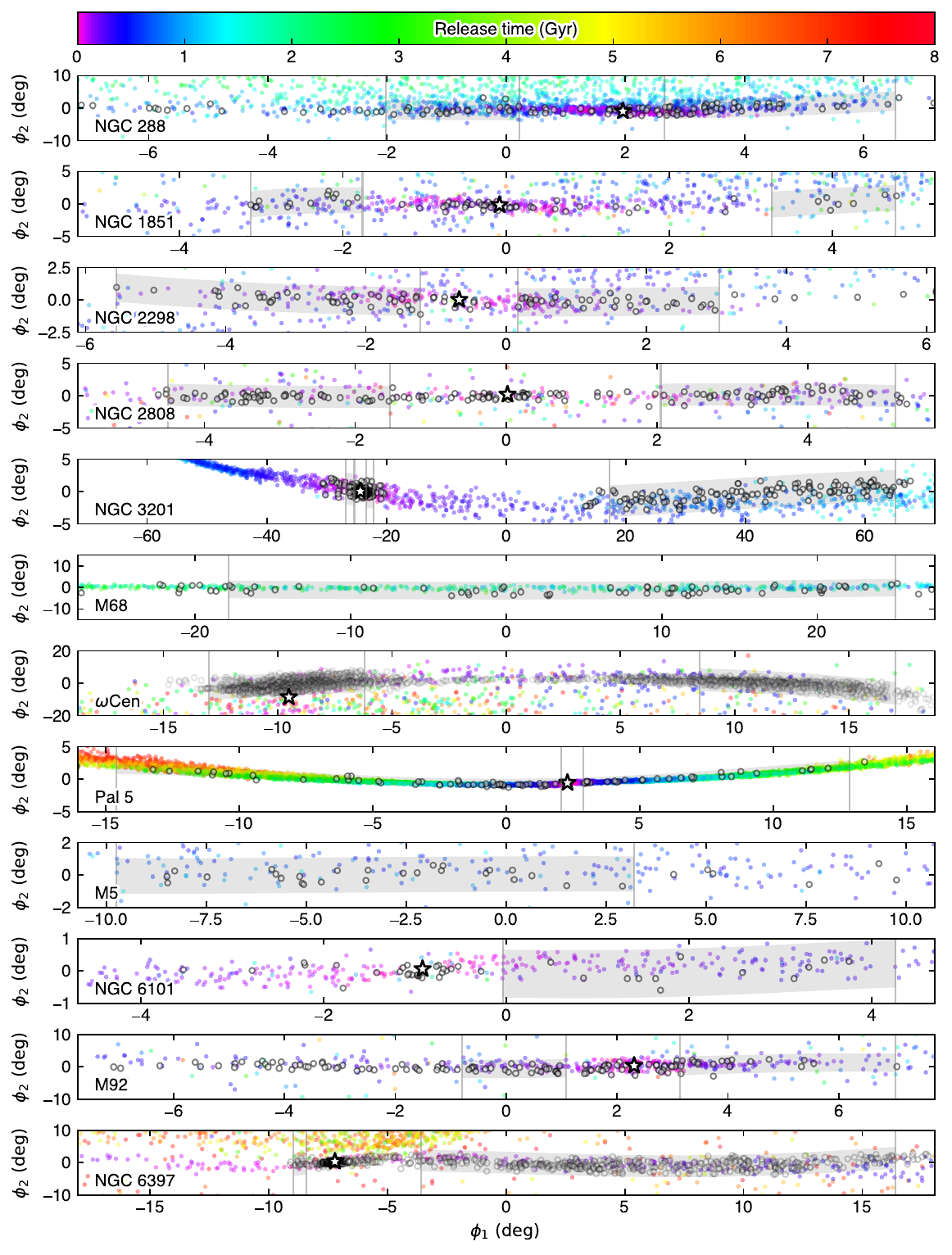}
    \vspace{-3mm}
    \caption{\textit{Gaia} observations of 12 GC streams (black open symbols) and their corresponding mock streams (colored solid symbols). Stream segments are represented as gray regions. The widths of these regions stand for $\pm3\,\sigma_{\rm detect}$. The mock stream particles are color-coded by their lookback release time, as indicated by the colorbar. The progenitor clusters are shown as star symbols.}
    \label{fig:stream_fit}
\end{figure*}

The \citetalias{ibata_charting_2024} catalog applies the \texttt{STREAMFINDER} algorithm to select stream stars. The \texttt{STREAMFINDER} algorithm assumes that each stream segment is a tube with a Gaussian dispersion around its track, and \citetalias{ibata_charting_2024} specifically focused on thin streams by using a Gaussian width of only 50~pc. Consequently, stars in the outskirts of wide streams (e.g., $\omega$Cen) are likely excluded from the catalog. Additionally, streams that cross the Galactic disk (e.g., NGC 3201) may be contaminated by disk stars, causing \texttt{STREAMFINDER} to miss parts of these streams. If we define the probability of stream stars being identified by \citetalias{ibata_charting_2024} as the spatial selection function $f_{\rm sel}(\phi_1, \hat{\phi}_2)$ of sky coordinates, this function tends to decrease towards the outskirts of streams and in regions with high contamination. However, this selection function is not explicitly provided in \citetalias{ibata_charting_2024}. Here, we estimate $f_{\rm sel}$ and refer to our estimate as $\tilde{f}_{\rm sel}$.

First, we assume that the selection function can be decomposed into separate dependencies on $\phi_1$ and $\hat{\phi}_2$: $\tilde{f}_{\rm sel}(\phi_1, \hat{\phi}_2) = \tilde{f}_{{\rm sel},1}(\phi_1) \tilde{f}_{{\rm sel},2}(\hat{\phi}_2)$. To estimate the $\hat{\phi}_2$-dependence for each stream, we assume that the detected stars follow a Gaussian distribution, $\hat{\phi}_{2, {\rm detect}} \sim \mathcal{N}(0, \sigma_{\rm detect}^2)$, where $\sigma_{\rm detect}$ is the standard deviation of $\hat{\phi}_2$ for all detected stars in the stream. Similarly, we assume that all stream stars follow a Gaussian distribution, approximated by our mock stream, $\hat{\phi}_{2, {\rm all}} \sim \mathcal{N}(0, \sigma_{\rm mock}^2)$. We have
\begin{equation}
    \exp\left(-\frac{\hat{\phi}_2^2}{2\sigma_{\rm mock}^2}\right)\tilde{f}_{{\rm sel},2}(\hat{\phi}_2)=\exp\left(-\frac{\hat{\phi}_2^2}{2\sigma_{\rm detect}^2}\right)
\end{equation}
which yields
\begin{equation}
    \tilde{f}_{{\rm sel},2}(\hat{\phi}_2) = \exp\left( \frac{\hat{\phi}_2^2}{2\sigma_{\rm mock}^2} -\frac{\hat{\phi}_2^2}{2\sigma_{\rm detect}^2}\right).
    \label{eq:selection}
\end{equation}
We also assume that $\tilde{f}_{{\rm sel},2} = 1$ at $\hat{\phi}_2 = 0$. 

The $\phi_1$-dependence of $f_{\rm sel}$ can diverge dramatically in different regions of the stream. A stream is usually divided into several segments with high $\tilde{f}_{{\rm sel},1}$, separated by voids with low $\tilde{f}_{{\rm sel},1}$. To identify segments where $\tilde{f}_{{\rm sel},1}$ is high, we first calculate the 1D number density of stars/particles along $\phi_1$ for both the detected and mock streams (within $|\hat{\phi}_2|<3\sigma_{\rm detect}$) using Gaussian kernel density estimation. The bandwidth is determined using the method by \citet{scott_multivariate_2015}. We then normalize the mock stream to match the observed maximum and identify segment boundaries where the observed density is higher or lower than the mock density by 50\%. The observed density below this threshold likely reflects low completeness, while a mismatch between the mock and observed streams may lead to observed density either below or above this threshold. However, choosing different thresholds between $25-75\%$ only weakly influences our results.

We set $\tilde{f}_{{\rm sel},1}=1$ within the identified segments and zero otherwise. We also exclude stars with $\Delta\phi_1$ within the tidal radius of the progenitor. Detection of these stars may be highly contaminated by the GC. This splits most streams into the trailing and leading segments separated by at least two tidal radii. 

Finally, by inserting our estimates for $M_{\rm sel}$ and $f_{\rm sel}$ back to Eq.~(\ref{eq:mass_loss}), we obtain the expression of $\dot{M}$ using the tracer particle number loss rate $\dot{N}_{\rm tracer}$,
\begin{equation}
    \dot{M}\approx\frac{\dot{N}_{\rm tracer}\sum_{k=1}^{N_{\rm\star,obs}} m_k \, w_k}{\sum_j \tilde{f}_{\rm sel}(\phi_{1,j},\hat{\phi}_{2,j})}
    \label{eq:mass_loss}
\end{equation}
where the summation in the denominator is over tracer particles in the mock stream, while the summation in the numerator is over stars above the detection limit in the \citetalias{ibata_charting_2024} catalog.

\section{Results}
\label{sec:results}

We generate mocks for 12 of the 16 streams associated with GCs from \citetalias{ibata_charting_2024}, excluding three streams for which we fail to reproduce morphology and one stream with too few stars (NGC 7089). We release tracer particles with a uniform rate $\dot{N}_{\rm tracer} = 1\ {\rm Myr^{-1}}$. In Fig.~\ref{fig:stream_fit}, we show the distribution of detected stream stars and mock particles in the great circle frame.

Although our mock streams match the morphology of most detected streams, some mock streams tend to be significantly wider than their detected counterparts (e.g., NGC 288). However, the newly-released portion of the mock streams can still match the width of the detected streams, while the older portion is noticeably wider. This suggests that the detected streams were released more recently than our integration duration of 8~Gyr. As shown in Appendix~\ref{sec:duration}, most streams formed within the last 1~Gyr.

Yet, since we weight the mock stream particles by the spatial selection function, we effectively focus our analysis on the region where the mock and detected streams overlap. Since the mismatched regions are down-weighted by the selection function, the early-released mock particles do not significantly affect our estimation of the mass loss rate.

Next, we compute the mass loss rate for each stream by matching the mass densities of mock and detected streams. For streams with multiple segments, we calculate the mass loss rate using all segments combined. The uncertainty $\sigma_{\log\dot{M}}$ is dominated by two sources. The first is the error due to the small number of observed stars. Through Monte Carlo sampling of the MF from a fixed $M_{\rm sel}$, we find that the number of observable stars roughly follows the Poisson's distribution. The Poisson's error can reach up to $\sim 0.1$~dex for streams with fewer observable stars (e.g., NGC 6101). The second source is propagated from the uncertain stellar mass function when correcting for the missing faint stars. This uncertainty is typically $0.1$~dex (see Appendix~\ref{sec:mf}). We calculate the total uncertainty by taking the Euclidean norm of these two sources. We also calculate the number loss rates $\dot{N}$ by dividing the mass loss rates by the mean stellar mass, averaged over the MF. In Table~\ref{tab:properties}, we list the estimated mass/number loss rates and their associated uncertainties.

We also verify that our mass loss rates are typically higher than the theoretical evaporation rates \citep[$\dot{M}_{\rm evap}=M_{\rm GC}/t_{\rm evap}$, with $t_{\rm evap} \equiv f t_{\rm rh}$, see Eqs.~7.108 and 7.141 in][using $f=20$]{binney_galactic_2008} by a factor of $\lesssim 2$, indicating that while evaporation is an important source of mass loss, it underestimates the total mass loss in a tidal field. The only exceptions are $\omega$Cen, Pal 5, and NGC 6101, where our rates are significantly higher than the evaporation rates by a factor of $10-30$. This is likely because these GCs are either very close to the Galactic center ($\omega$Cen) or have half-mass radii that are very close to their tidal radii (Pal 5 and NGC 6101).

To verify our method, we also perform a similar analysis for Pal 5 from a different dataset in Appendix~\ref{sec:cfht}. Remarkably, our method yields results consistent with those from \textit{Gaia}. This consistency supports the reliability of our correction for missing stars and our reconstruction of the selection function.

\begin{figure*}
    \centering
    \includegraphics[width=\linewidth]{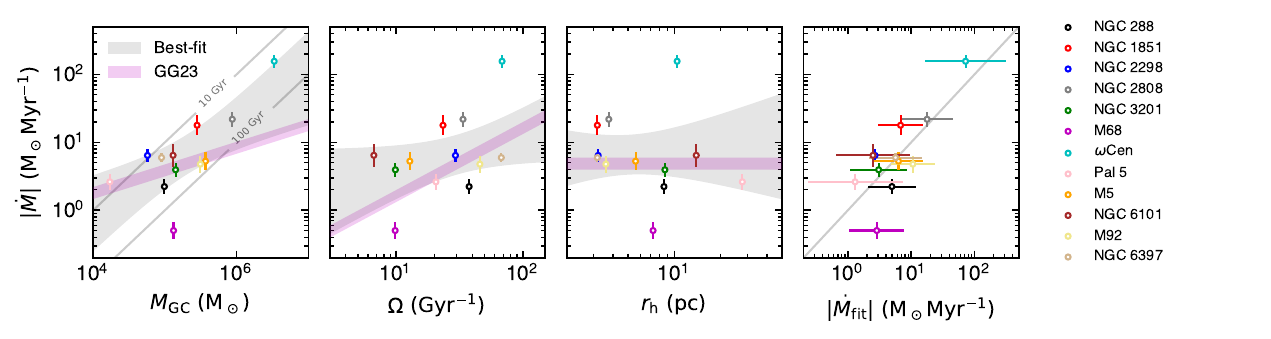}
    \vspace{-7mm}
    \caption{Mass loss rate of 12 GCs against $M_{\rm GC}$ (\textit{first panel}), $\Omega$ (\textit{second panel}), $r_{\rm h}$ (\textit{third panel}), and best-fit mass loss rate $\dot{M}_{\rm fit}$ (\textit{fourth panel}). The uncertainty of $\dot{M}$ (from Poisson's error and MF slope) is plotted as vertical errorbars. Horizontal errorbars in the \textit{fourth panel} represent the uncertainty of the best-fit power-law relation (Eqs.~\ref{eq:powerlaw} and \ref{eq:bestfit}). Gray shaded regions in the first three panels show the best-fit mass loss rate and the associated uncertainty, with the other two GC properties set to the reference values: $M_{\rm GC}=2\times10^5\Msun$, $\Omega=30\ {\rm Gyr^{-1}}$, and $r_{\rm h}=5$~pc. Magenta regions show the $(a,b,c)=(1/3,1,0)$ model from \citet{gieles_mass-loss_2023} with $|\dot{M}_{\rm ref}|=30-45\Msun\,{\rm Myr^{-1}}$. Gray lines in the \textit{first panel} indicate the expected lifetime if GCs continue to dissolve at the current rates. Gray line in the \textit{fourth panel} indicates one-to-one agreement.}
    \label{fig:mdot_scaling}
\end{figure*}

In Fig.~\ref{fig:mdot_scaling}, we plot $\dot{M}$ against three GC properties: mass $M_{\rm GC}$, orbital frequency $\Omega$, and half-mass radius $r_{\rm h}$. The values for $M_{\rm GC}$ and $r_{\rm h}$ are taken directly from the fourth edition of the \citet{hilker_galactic_2019} catalog. We calculate the orbital frequency as $\Omega = 2\pi/T$, where the orbital period $T$ is the time interval between two adjacent pericenter passages.

\begin{deluxetable*}{cccccccccc}
\tablecaption{Summary of GC and stream properties. $M_{\rm GC}$, $r_{\rm h}$, $\alpha$, and the distance to the cluster $d$ are directly taken from the \citet{hilker_galactic_2019} catalog or \citet[][for Pal 5 only]{erkal_sharper_2017}. $\Omega$, $|\dot{M}|$, $|\dot{N}|$, $|\Delta\phi_{\rm 1,max}|$, and duration are derived in this work.}
\label{tab:properties}
\tablehead{
\colhead{GC} & \colhead{$M_{\rm GC}$} & \colhead{$\Omega$} & \colhead{$r_{\rm h}$} & \colhead{$\alpha$} & \colhead{$d$} & \colhead{$|\dot{M}|$} & \colhead{$|\dot{N}|$} & \colhead{$|\Delta\phi_{\rm 1,max}|$} & \colhead{Duration} \\
 & $(10^5\Msun)$ & $({\rm Gyr^{-1}})$ & $({\rm pc})$ &  & $({\rm kpc})$ & $({\rm M_\odot\,Myr^{-1}})$ & $({\rm Myr^{-1}})$ & $({\rm deg})$ & $({\rm Gyr})$ 
}
\startdata
NGC 288 & 0.98 & 37.6 & 8.6 & 0.66 & 9.0 & $2.2 \pm 0.5$ & $6.2 \pm 1.4$ & 4.6 & $0.70 \pm 0.37$ \\
NGC 1851 & 2.81 & 23.4 & 3.2 & 0.70 & 11.9 & $18 \pm 5$ & $50 \pm 15$ & 4.9 & $0.45 \pm 0.29$ \\
NGC 2298 & 0.57 & 29.4 & 3.2 & -0.17 & 9.8 & $6.4 \pm 1.1$ & $14 \pm 2$ & 4.9 & $0.22 \pm 0.15$ \\
NGC 2808 & 8.69 & 33.7 & 3.8 & 0.60 & 10.1 & $22 \pm 5$ & $59 \pm 13$ & 5.1 & $0.13 \pm 0.11$ \\
NGC 3201 & 1.41 & 9.9 & 8.7 & 1.02 & 4.7 & $3.9 \pm 0.8$ & $13 \pm 3$ & 89.5 & $1.1 \pm 0.4$ \\
M68 & 1.32 & 9.8 & 7.3 & 1.17 & 10.4 & $0.50 \pm 0.13$ & $1.7 \pm 0.5$ & 86.1 & $2.3 \pm 0.8$ \\
$\omega$Cen & 33.40 & 68.5 & 10.4 & 0.80 & 5.4 & $160 \pm 30$ & $450 \pm 80$ & 26.6 & $0.42 \pm 0.31$ \\
Pal 5 & 0.17 & 20.6 & 27.6 & 0.84 & 23.6 & $2.7 \pm 0.6$ & $8.1 \pm 1.9$ & 16.9 & $6.2 \pm 1.7$ \\
M5 & 3.68 & 12.8 & 5.6 & 0.76 & 7.5 & $5.3 \pm 1.4$ & $15 \pm 4$ & 40.1 & $0.64 \pm 0.24$ \\
NGC 6101 & 1.30 & 6.7 & 13.9 & 0.90 & 14.4 & $6.4 \pm 2.3$ & $20 \pm 7$ & 5.2 & $0.30 \pm 0.11$ \\
M92 & 3.12 & 45.8 & 3.6 & 0.83 & 8.5 & $4.8 \pm 1.2$ & $14 \pm 4$ & 4.7 & $0.24 \pm 0.16$ \\
NGC 6397 & 0.90 & 67.5 & 3.2 & 0.32 & 2.5 & $6.0 \pm 0.7$ & $15 \pm 2$ & 23.6 & $0.31 \pm 0.16$ \\
\enddata
\end{deluxetable*}

We observe positive correlation between $\dot{M}$ and $M_{\rm GC}$. If these GCs continue to lose mass at their current rates, most are expected to dissolve within $10$ to $100$~Gyr. Notably, Pal 5 and NGC 2298 have relatively high rates for their low masses, leading to dissolution times $<10$~Gyr. There is also positive correlation between $\dot{M}$ and $\Omega$, though with significant scatter. In contrast, $\dot{M}$ shows only weak correlation with $r_{\rm h}$. 

To quantitatively study the scaling relation between $\dot{M}$ and the three GC properties, we fit our data using the following multivariate power-law function:
\begin{equation}
    \dot{M}_{\rm fit} = \dot{M}_{\rm ref}\left(\frac{M_{\rm GC}}{2\times10^5\Msun}\right)^a\left(\frac{\Omega}{30\ {\rm Gyr^{-1}}}\right)^b\left(\frac{r_{\rm h}}{5\ {\rm pc}}\right)^c.
    \label{eq:powerlaw}
\end{equation}
We define the likelihood of the data given the model as:
\begin{equation}
    {\cal L} = \prod_i \frac{1}{\sigma_i\sqrt{2\pi}}\exp\left(-\frac{\Delta_i^2}{2\sigma_i^2}\right)
    \label{eq:likelihood}
\end{equation}
where $\Delta_i \equiv \log_{10}|\dot{M}_i| - \log_{10}|\dot{M}_{\rm fit}|$, and $\sigma_i^2 \equiv \sigma_{\log\dot{M}}^2 + \sigma_{\rm int}^2$ represents the total logarithmic uncertainty. Here, we model the intrinsic scatter of $\dot{M}$ as a log-normal distribution with standard deviation $\sigma_{\rm int}$.

The best-fit parameters are obtained by maximizing the likelihood function. To estimate the uncertainty of these parameters, we resample our data 1,000 times using the bootstrapping technique, yielding the following results:
\begin{equation}
    \begin{array}{rl}
        |\dot{M}_{\rm ref}| &= 6.6^{+2.8}_{-2.0}\Msun\,{\rm Myr^{-1}} \\
        \sigma_{\rm int} &= 0.25\pm0.10\ {\rm dex} \\
        a &= 0.66\pm0.37 \\
        b &= 0.54\pm0.55 \\
        c &= 0.12\pm0.72
    \end{array}
    \label{eq:bestfit}
\end{equation}
with weak correlation between each pair of the parameters. We compare $\dot{M}_{\rm fit}$ with these parameters to the actual $\dot{M}$ in the \textit{fourth panel} of Fig.~\ref{fig:mdot_scaling}. The two rates agree within approximately one standard deviation, indicating that our best-fit provides a reliable estimate of the GC mass loss rate.

We note that $c$ is consistent with being zero with large scatter, indicating that $r_{\rm h}$ does not significantly influence the mass loss rate. Although $a$ and $b$ have considerable scatter, both are above zero by $1–2$ times the standard deviation. We also perform a one-parameter fit with fixed values $b=1$ and $c=0$. In this case, the slope of the mass dependence is reduced: $a=0.58\pm 0.19$ ($b=1$, $c=0$).

Compared to scaling relations derived from N-body simulations, such as those from \citet[][also shown in Fig.~\ref{fig:mdot_scaling}]{gieles_mass-loss_2023}, our power-law slopes $(a, b, c)$ are consistent with their no-BH model, $(a, b, c) = (1/3, 1, 0)$, within one standard deviation.
For a reference GC with $M_{\rm GC} = 2 \times 10^5\ \Msun$, $\Omega = 30\ {\rm Gyr^{-1}}$, and $r_{\rm h} = 5\ {\rm pc}$, our relation gives $|\dot{M}| = 6.6^{+2.8}_{-2.0}\ \Msun\,{\rm Myr^{-1}}$. This is slightly higher than $|\dot{M}| = 4 - 6\ \Msun\,{\rm Myr^{-1}}$ from the models in \citet{gieles_mass-loss_2023}\footnote{Note that \citet{gieles_mass-loss_2023} used the tidal frequency $\Omega_{\rm tid}$ instead of the orbital frequency $\Omega$. The former is $\sqrt{2}$ times larger than $\Omega$ in their Galactic potential, which we have accounted for in our comparison.}, but still within $1\sigma$.

It is worth noting that previous theoretical works have reported different mass loss rates for Pal 5. \citet{gieles_supra-massive_2021} suggested an average mass loss rate $M_{\rm dot} \approx 10~\Msun\,{\rm Myr^{-1}}$, assuming Pal 5 is rich in black holes (BHs), whereas \citet{kupper_globular_2015} estimated $M_{\rm dot} = 7.9_{-6.4}^{+5.3}~\Msun\,{\rm Myr^{-1}}$, which is consistent with our measurement. However, this discrepancy does not necessarily rule out the results of \citet{gieles_supra-massive_2021}, as our particle spray method is not calibrated for BH-rich GCs. Additional observations beyond stream density are required to disentangle the degeneracy between BH-rich and BH-poor scenarios.

\section{Summary and discussion}
\label{sec:summary}

We present the first catalog of directly observed mass loss rates for Galactic GCs by comparing the densities of detected and mock streams. Using the state-of-the-art particle spray algorithm by \citetalias{chen_improved_2025}, we generate mock tidal streams around GCs with associated streams detected by \citetalias{ibata_charting_2024} in \textit{Gaia} DR3. Our robust pipeline estimates the mass loss rate $\dot{M}$ accounting for missing faint stars below the detection limit and the spatial selection function along and perpendicular to the stream track.

We successfully reproduce stream morphology around 12 GCs (Fig.~\ref{fig:stream_fit}). This allows us to make precise measurement of their mass loss rate, listed in Table~\ref{tab:properties}. We also examine correlations between $\dot{M}$ and various GC properties (Fig.~\ref{fig:mdot_scaling}). While we find no statistically significant trend with $r_{\rm h}$, we observe positive correlations between $\dot{M}$ and both $M_{\rm GC}$ and $\Omega$. When parameterized by power-law functions, our best-fit parameters are consistent with those from N-body simulations by \citet{gieles_mass-loss_2023}.

We emphasize that this work only considers the mass loss of main-sequence stars due to tidal stripping. More violent dynamical processes, such as binary kicks, can eject stars at speeds far exceeding the escape velocity. These stars, which contribute $10-20\%$ of the total mass loss \citep{weatherford_stellar_2023}, may not end up in the stream and are therefore not captured by our method. Additionally, we do not account for mass loss due to stellar evolution, which dominates during the initial phase after cluster formation but becomes negligible over the timescales relevant to streams. Furthermore, our correction for missing stellar mass below the detection limit (\S2.4) does not include dark remnants such as BHs. While this could significantly affect BH-rich GCs such as Pal 5, the typical BH ejection rate is around $0.01~{\rm Myr^{-1}}$ or $0.2~\Msun\,{\rm Myr^{-1}}$ \citep{weatherford_stellar_2023}, which is small for most GCs in this work. Therefore, while our method captures the main source of mass loss, it may slightly underestimate the total mass loss rate by several tens of percent.

However, \textit{Gaia} data remain largely incomplete for most streams due to their low surface brightness and entangled morphology. Future instruments could significantly improve the accuracy of GC mass loss rate estimates. For instance, the \textit{Vera C. Rubin Observatory} and the \textit{Nancy Grace Roman Space Telescope} have the potential not only to detect up to $\sim1000$ GC streams \citep{pearson_forecasting_2024}, but also to better constrain mass loss rates through wide-field, deep, homogeneous, and complete photometric surveys.

Additionally, most stream stars still lack full 6D phase space information, especially the line-of-sight velocities. This restricts our analysis to streams associated with known GCs, where we can generate mock streams using the positions and velocities of their progenitor clusters. Currently operating facilities, such as the Dark Energy Spectroscopic Instrument \citep{desi_collaboration_overview_2022}, could efficiently obtain line-of-sight velocities with high accuracy. This would allow us to extend our analysis to streams originating from fully disrupted GCs \citep[e.g., GD-1,][]{valluri_gd-1_2024}.

% \newpage
\section*{Acknowledgements}
We thank Eric Bell, Monica Valluri, Ana Bonaca, Eugene Vasiliev, and Sergey Koposov for insightful discussions. We also thank the reviewer for their suggestions and comments, which have improved the quality of this work. YC and OG were supported in part by the U.S. National Science Foundation through grant AST-1909063 and by National Aeronautics and Space Administration through contract NAS5-26555 for Space Telescope Science Institute program HST-AR-16614. HL is supported by the National Key R\&D Program of China No. 2023YFB3002502, the National Natural Science Foundation of China under No. 12373006, and the China Manned Space Program through its Space Application System. This research benefited from the Dwarf Galaxies, Star Clusters, and Streams Workshop hosted by the Kavli Institute for Cosmological Physics.

\software{
\texttt{agama} \citep{vasiliev_agama_2019}, 
\texttt{numpy} \citep{harris_array_2020}, 
\texttt{matplotlib} \citep{hunter_matplotlib_2007}, 
\texttt{scipy} \citep{virtanen_scipy_2020}, 
\texttt{astropy} \citep{the_astropy_collaboration_astropy_2018}, 
\texttt{gala} \citep{price-whelan_gala_2017,price-whelan_adrngala_2024}, 
\texttt{pandas} \citep{the_pandas_development_team_pandas-devpandas_2024}
}

% \newpage
\appendix
\vspace{-6mm}

\section{Duration of streams}
\label{sec:duration}

The release time of the mock streams offers a unique opportunity to measure the time required to populate the observed streams, i.e., the duration of the streams. Here, we rigorously define and calculate the duration of each stream.

We fit the release time of the mock stream particles as a linear function of $\Delta\phi_1$ using a likelihood-maximization methodm, where $\Delta\phi_1\equiv\phi_1-\phi_{\rm 1,GC}$ is the angular distance to the progenitor GC. We also include intrinsic scatter, which is modeled as a linear function of release time. Since particles in the trailing and leading arms move in opposite directions, we independently fit the $\Delta\phi_1$–-release time relation for each arm. We limit the fit to regions with $|\hat{\phi}_2| < 3\sigma_{\rm detect}$ to exclude early-released particles that have wrapped around the orbit.

For most streams, we find significant correlation between $\Delta\phi_1$ and the release time. The ratio of the two yields the diffusion rate, which ranges from $3^\circ\,{\rm Gyr^{-1}}$ (Pal 5) to $100^\circ\,{\rm Gyr^{-1}}$ (NGC 3201). Multiplying the diffusion rate by the distance to the GC yields the physical diffusion velocity. We find that the diffusion velocity is generally consistent with the escape velocity at the GC's current tidal radius, $v_{\rm esc}\equiv\sqrt{2GM_{\rm GC}/r_{\rm tid}}$. The ratio between the two velocities ranges from 0.34 for streams near the apocenter (NGC 288) to 1.76 for those near the pericenter (NGC 3201).

We define the duration of a stream as the release time at the boundary of the stream segment. If a stream has multiple segments, we define the duration at the farthest boundary $|\Delta\phi_{\rm 1,max}|$. This ensures that the stream can extend to all detected segments within the calculated duration.

The only exception is NGC 288, which has an eccentric orbit ($e \approx 0.7$) and is currently near its apocenter below the Galactic plane. As a result, both the trailing and leading arms curve toward the same side of the progenitor. NGC 288 is also dynamically hot, so that the two arms are strongly mixed in $\phi_1$, making it difficult to detect a clear correlation between release time and $\phi_1$ (see Fig.~\ref{fig:stream_fit}). For this stream, we define the duration as the average release time of particles within $|\hat{\phi}_2| < 3\sigma_{\rm detect}$.

We list $\Delta\phi_{\rm 1,max}$ and the duration in Table~\ref{tab:properties}. Most streams are dynamically young, with duration $<1$~Gyr. Although Pal 5 is not the longest stream, it has a significantly longer duration of approximately 6~Gyr due to its low diffusion rate. We emphasize that the duration refers to the time required to form the detectable portion of the stream. Therefore, we should not over-interpret Pal 5's long duration, as other streams may also have longer durations but are not yet detected by \citetalias{ibata_charting_2024}.

\section{Dependence on MF}
\label{sec:mf}

\begin{figure*}
    \centering
    \includegraphics[width=0.8\linewidth]{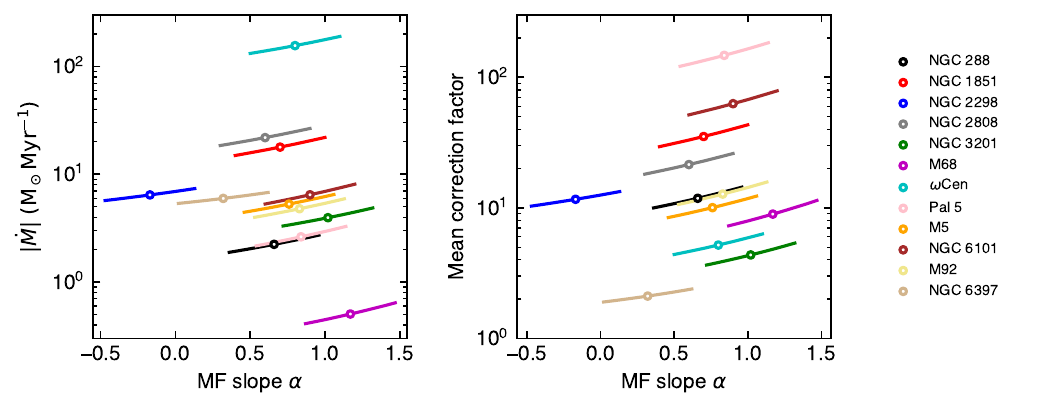}
    \vspace{-3mm}
    \caption{Mass loss rate (\textit{left panel}) and mean correction factor (as in Eq.~(\ref{eq:correction}), \textit{right panel}) of 12 GCs derived from different MF slopes $\alpha$. For each GC, we plot the range around the central value with $\Delta\alpha=\pm0.3$, which is used in this work as the dispersion of $\alpha$.}
    \label{fig:alpha_test}
\end{figure*}

As mentioned in \S\ref{sec:obs_mass}, we assume the stellar MF follows a power law $\psi(m) \propto m^{-\alpha}$. The MF is used to correct for the missing stellar mass via Eq.~(\ref{eq:correction}). Since most stars fall below the minimum observable mass, the correction factor is typically much greater than unity and can be sensitive to the choice of the MF slope $\alpha$. Typically, $w_i$ ranges from 2 to 200 depending on the distance to the stream. In Fig.~\ref{fig:alpha_test} we examine a wide range of $\alpha$ from $-0.5$ to $1.5$ and study their impact on the inferred mass loss rate $\dot{M}$ and the mean correction factor. We find that $|\dot{M}|$ increases by approximately 0.1~dex for every 0.3 increment in $\alpha$. Although not visually obvious, closer streams like NGC 3201 ($d \approx 5$~kpc) and NGC 6397 ($d \approx 2$~kpc) are less sensitive to changes in $\alpha$, while more distant streams at $d = 10-20$~kpc tend to be more sensitive. This is because we can observe much fainter stars below the main sequence turnover for the closer streams. As a result, these closer streams also have smaller mean correction factors.

Although \citet{baumgardt_evidence_2023} reported an uncertainty of $\Delta\alpha = 0.1 - 0.2$ for their $\alpha$ measurements, we notice deviations among different studies of up to $0.5$ \citep{de_marchi_why_2007,paust_acs_2010,ebrahimi_new_2020,dickson_multimass_2023}. Therefore, we adopt a more conservative uncertainty of $\Delta\alpha = 0.3$ throughout this work. This translates to an uncertainty in $|\dot{M}|$ of about $0.1$~dex.

\section{Proof of the unbiased estimate of observed mass}
\label{sec:proof}

In Eq.~(\ref{eq:estimate}), $N_{\star}$ represents the total number of stars. By definition,
\begin{equation*}
    N_{\star}\equiv\iint f_{\rm sel}(\phi_1,\hat{\phi}_2)\Sigma_{N_{\star}}(\phi_1,\hat{\phi}_2) d\phi_1 d\hat{\phi}_2.
\end{equation*}
Given the mean stellar mass of the stellar population,
\begin{equation*}
    \left<m\right> \equiv \frac{\int_{m_{\rm limit}}^{m_{\rm max}}m\psi(m)dm}{\int_{m_{\rm limit}}^{m_{\rm max}}\psi(m)dm}.
\end{equation*}
we have $\Sigma(\phi_1, \hat{\phi}_2) = \left<m\right> \Sigma_{N_{\star}}(\phi_1, \hat{\phi}_2)$. Combining this with Eq.~(\ref{eq:m_obs}) gives $M_{\rm sel} = N_{\star} \left<m\right>$. Therefore, we can demonstrate that Eq.~(\ref{eq:estimate}) provides an unbiased estimate of $M_{\rm sel}$:
\begin{align*}
    \mathbb{E}[\tilde{M}_{\rm sel}]
    &= \sum_{i=1}^{N_{\star}} \mathbb{E}[\theta(m_i-m_{{\rm min},i})m_i]w_i \\
    &\equiv \sum_{i=1}^{N_{\star}} \frac{\int_{m_{{\rm min},i}}^{m_{\rm max}}m\psi(m)dm}{\int_{m_{\rm limit}}^{m_{\rm max}}\psi(m)dm}
        \frac{\int_{m_{\rm limit}}^{m_{\rm max}}m\psi(m)dm}{\int_{m_{{\rm min},i}}^{m_{\rm max}}m\psi(m)dm} \\
    &= \sum_{i=1}^{N_{\star}} \frac{\int_{m_{\rm limit}}^{m_{\rm max}}m\psi(m)dm}{\int_{m_{\rm limit}}^{m_{\rm max}}\psi(m)dm} 
    \equiv \sum_{i=1}^{N_{\star}}\left<m\right> \\
    &= N_{\star}\left<m\right> 
    = M_{\rm sel}.
\end{align*}

\section{Comparison with other data}
\label{sec:cfht}

In \S\ref{sec:obs_mass}, we introduced the correction factor in Eq.~(\ref{eq:correction}) to account for the missing stellar mass below the detection limit of \textit{Gaia} DR3. In \S\ref{sec:selection_function}, we reconstructed the spatial selection function of \citetalias{ibata_charting_2024}, assuming that $\hat{\phi}_2$ follows a Gaussian distribution. However, since \textit{Gaia} is only complete up to $G \approx 19$, the majority of stellar mass is missed, resulting in large and uncertain correction factors. Additionally, the accuracy of our reconstructed selection function is uncertain since the actual selection function is not explicitly provided in \citetalias{ibata_charting_2024}. To validate our corrections and assumptions, we apply our method to a different dataset with a deeper detection limit and a simpler selection function, and compare the inferred $\dot{M}$ with the results from \citetalias{ibata_charting_2024}.

\begin{figure*}
    \centering
    \includegraphics[width=0.345\linewidth]{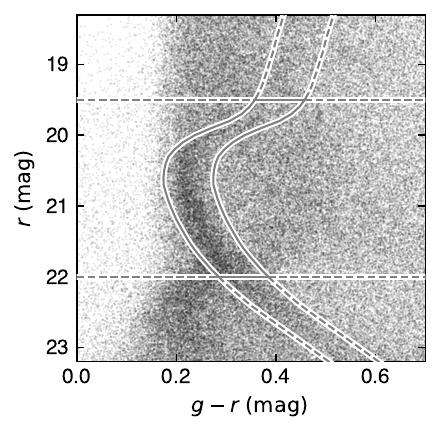}
    \includegraphics[width=0.645\linewidth]{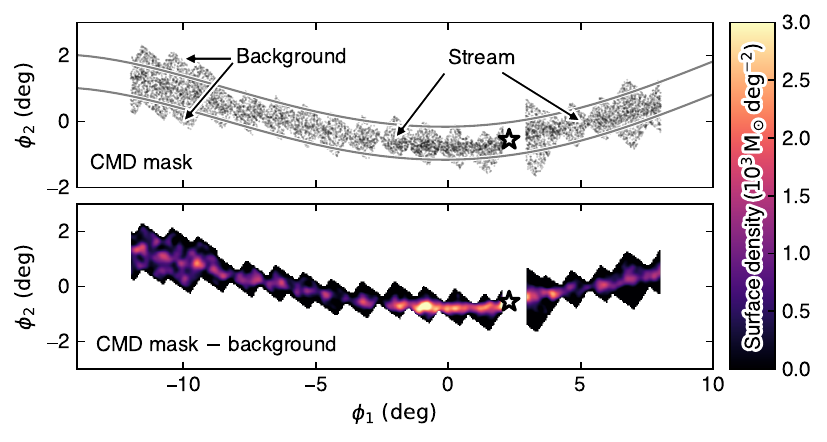}
    \vspace{-3mm}
    \caption{\textit{Left panel}: CFHT data in the extinction corrected $g-r$ color vs. $r$ magnitude diagram. The CMD mask for Pal 5 is shown as gray solid curves. \textit{Top right panel}: Selected Pal 5 stream stars in the great circle frame. We compute the background surface density in the outskirt regions with $|\hat{\phi}_2|>0.5^\circ$ (indicated by gray curves). The Pal 5 cluster is shown as the star symbol. \textit{Bottom right panel}: Surface density of the Pal 5 stream with the background subtracted. We use Gaussian kernel density estimation with $0.1^\circ$ bandwidth to estimate the density.}
    \label{fig:cfht}
\end{figure*}

We focus specifically on the Pal 5 stream, for which comprehensive photometric data are available from the Canada-France-Hawaii Telescope \citep[CFHT,][]{ibata_feeling_2016}. In the \textit{left panel} of Fig.~\ref{fig:cfht}, we show the CFHT data on an extinction-corrected color–magnitude diagram (CMD). This dataset is complete up to $r \approx 23$, which extends well below the main sequence turnover of the Pal 5 stars at $r = 20 - 21$. In contrast, \textit{Gaia} DR3 only barely reveals the turnover. For the following analysis, we set $r_{\rm limit} = 22$. Since the CFHT dataset does not employ additional selection criteria, the selection function is simply uniform, i.e., $f_{\rm sel} = 1$ across the entire data footprint.

We apply a CMD mask to select stars belonging to the Pal 5 stream. The mask has a width of 0.1~mag in the $g-r$ direction, centered on the \texttt{PARSEC} isochrone fit to Pal 5. We have verified that varying the mask width between $0.06 - 0.2$~mag does not significantly affect the inferred $\dot{M}$, as long as we correctly subtract background stars. We set the magnitude limit in the faint end as $r_{\rm limit} = 22$. Additionally, we avoid the red giant branch by applying a magnitude cut in the bright end at $r = 19.5$. As shown in the \textit{left panel} of Fig.~\ref{fig:cfht}, the mask covers the overdensity of the Pal 5 stream in the CMD.

In the \textit{top right panel} of Fig.~\ref{fig:cfht}, we plot the masked stars in the same great circle frame as in Fig.~\ref{fig:stream_fit}. The CFHT footprint only covers the region $\phi_1 = -12^\circ$ to $8^\circ$, which is smaller than the stream segments detected by \citetalias{ibata_charting_2024}. Therefore, we restrict both the plot and our analysis to the intersection of this region with the stream segments from \citetalias{ibata_charting_2024}.

Although the stream is surrounded by background stars, we can still observe it within $|\hat{\phi}_2| < 0.2^\circ$ around the track. We select regions with $|\hat{\phi}_2| > 0.5^\circ$ to estimate the density of background stars. We have verified that varying this threshold has little effect on the background star density, as long as the selected region is sufficiently far from the stream.

Next, we weight the mass of each star using Eq.~(\ref{eq:correction}) to estimate the total mass and mass distribution of the stream. In the \textit{lower right panel} of Fig.~\ref{fig:cfht}, we show the surface mass density of the stream, using Gaussian kernel density estimation with the background subtracted. 

Finally, we can calculate the mass loss rate using Eq.~(\ref{eq:mass_loss}). We obtain $\dot{M} = 2.63\ {\rm M_\odot\,Myr^{-1}}$ from the CFHT dataset, which closely matches the value derived from \textit{Gaia} DR3, $\dot{M} = 2.7 \pm 0.6\ {\rm M_\odot\,Myr^{-1}}$. Since the CFHT data is significantly deeper than \textit{Gaia}, the correction factor is only $\sim 6$ for CFHT, compared to $>100$ for \textit{Gaia}. This consistency supports that we successfully recover the missing mass between the detection limits of \textit{Gaia} and CFHT, as well as reconstruct the selection function for the \citetalias{ibata_charting_2024} catalog.

\bibliography{references}
\bibliographystyle{aasjournal}

\end{document}